\documentclass[11pt]{article}          

\usepackage{graphicx}
\usepackage[dvips]{color}

\newcommand\Softitle[1]{\Large \bf \noindent \begin{center} #1
\end{center}\rm \normalsize \vskip.125in }%

\newcommand\Sofauthor[1]{\vskip.1in\noindent%
\large  \begin{center} \textsf{#1} \end{center}\rm \vskip-.2in}

\long\def\symbolfootnote[#1]#2{\begingroup%
\def\thefootnote{\fnsymbol{footnote}}\footnote[#1]{#2}\endgroup}

\let\title\Softitle
\let\author\Sofauthor

\let\address\Sofaddress

\begin{document}


%
%

\title{A mechanism giving a finite value to the speed of light, and some experimental consequences}

\author{Marcel Urban\footnote{urban@lal.in2p3.fr}, Fran\c cois Couchot, Xavier Sarazin}

\address{LAL, Univ Paris-Sud, CNRS/IN2P3, Orsay, France.}

\begin{abstract}
We admit that the vacuum is not empty but is filled with continuously appearing and disappearing virtual fermion pairs.
We show that if we simply model the propagation of the photon in vacuum as a series of transient captures within the virtual pairs, we can derive the finite light velocity $c$ as the average delay on the photon propagation. We then show that the vacuum permittivity $\epsilon_0$ and permeability $\mu_0$ originate from the polarization and the magnetization of the virtual fermions pairs.
Since the transit time of a photon is a statistical process within this model, we expect it to be fluctuating. 
We discuss experimental tests of this prediction. We also study vacuum saturation effects under high photon density conditions.
\end{abstract}

\section{Introduction}

The speed of light in vacuum $c$, the vacuum permittivity  $\epsilon_0$ and the vacuum permeability $\mu_0$ are widely considered as being fundamental constants and their values, escaping any physical explanation, are commonly assumed to be invariant in space and time. In this paper, we propose a mechanism based upon a "natural" quantum vacuum description which leads to sensible estimations of these three electromagnetic constants, and we start drawing some consequences of this perspective.

The idea that the vacuum is a major partner of our world is not new. It plays a part for instance in the Lamb shift~\cite{lamb}, the variation of the fine structure constant with energy~\cite{bare-charge}, and the electron~\cite{magnetic-moment} and muon~\cite{Davier} anomalous magnetic moments. But these effects, coming from the so-called vacuum polarization, are second order corrections.
Quantum Electrodynamics is a perturbative approach to the electromagnetic quantum vacuum. This paper is concerned with the description of the fundamental, non perturbed, vacuum state.

While we were writing this paper, Ref.~~\cite{Leuchs} proposed a similar approach to give a physical origin to $\epsilon_0$ and $\mu_0$. Although this derivation is different from the one we propose in this paper, the original idea is the same: {"The physical electromagnetic constants, whose numerical values are simply determined experimentally, could emerge naturally from the quantum theory"~\cite{Leuchs}}. 
We do not know of any other paper proposing a direct derivation of $\epsilon_0$ and $\mu_0$ or giving a mechanism based upon the quantum vacuum leading to $c$.

The most important consequence of our model is that $c$, $\epsilon_0$ and $\mu_0$ are not fundamental constants but are observable parameters of the quantum vacuum: they can vary if the vacuum properties vary in space or in time.

The paper is organized as follows. First we describe our model of the quantum vacuum filled with virtual charged fermion pairs and we show that, by modeling the propagation of the photon in this vacuum as a series of interactions with virtual pairs, we can derive its velocity. Then we show how $\epsilon_0$ and $\mu_0$ might originate from the electric polarization and the magnetization of these virtual pairs. Finally we present two experimental consequences that could be at variance with the standard views and in particular we predict statistical fluctuations of the transit time of photons across a fixed vacuum path.

%
%

\section{An effective description of quantum vacuum}
\label{sec:model}

The vacuum is assumed to be filled with virtual charged fermion pairs (particle-antiparticle). The other vacuum components are assumed not to be connected with light propagation (we do not consider intermediate bosons, nor supersymmetric particles). All known species of charged fermions are taken into account: the three families of charged leptons $e$, $\mu$ and $\tau$ and the three families of quarks ($u$, $d$), ($c$, $s$) and ($t$, $b$), including their three color states. This gives a total of $21$ pair species, noted $i$.

A virtual pair is assumed to be the product of the fusion of two virtual photons of the vacuum. Thus its total electric charge and total color are null,
and we suppose also that the spins of the two fermions of a pair are antiparallel. The only quantity which is not conserved is therefore the energy and this is, of course, the reason for the limited  lifetime of the pairs. We assume that first order properties can be deduced assuming that pairs are created with an average energy, not taking into account a full probability density of the pair kinetic energy. Likewise, we will neglect the total momentum of the pair.

We describe this vacuum in terms of five quantities for each pair species: average energy, lifetime, density, size of the pairs and cross section with photons.

We use the notation $Q_i = q_i/e$, where $q_i$ is the modulus of the $i$-kind fermion electric charge and $e$ the modulus of the electron charge. 

\textbullet \ The average energy  $W_i$ of a pair is taken proportional to its rest mass energy $2W_i^0$, where $W_i^0$ is the fermion $i$ rest mass energy: 
\begin{eqnarray}
\label{eq:energy}
W_i\  = K_W\ 2 W_i^0 ,
\end{eqnarray}
where $K_W$ is an unknown constant, assumed to be independent from the fermion type. We take $K_W$ as a free parameter, greater than unity. The value of $K_W$ could be calculated if we knew the energy spectrum of the virtual photons together with their probability to create virtual pairs.

\textbullet \ The pair  lifetime $\tau_i$ follows from the Heisenberg uncertainty principle \newline $(W_i\tau_i=\hbar/2)$. So
\begin{eqnarray}
\label{eq:tau}
\tau_i = \frac{1}{K_W}\frac{\hbar}{4 W_i^0} .
\end{eqnarray}

\textbullet \ We assume that the virtual pair densities $N_i$ are driven by the Pauli Exclusion Principle. Two pairs containing two identical virtual fermions in the same spin state cannot show up at the same time at the same place. However at a given location we may find 21 pairs since different fermions can superpose spatially. In solid state physics the successful determination of Fermi energies~\cite{Kittel} implies that one electron spin state occupies a hyper volume $h^3$. We assume that concerning the Pauli principle, the virtual fermions are similar to the real ones. Noting $\Delta x_i$ the spacing between identical virtual $i-$type fermions and $p_i$ their average momentum, the one dimension hyper volume is $p_i\Delta x_i$ and dividing by $h$ should give the number of states which we take as one per spin degree of freedom. The relation between $p_i$ and $\Delta x_i$ reads $p_i\Delta x_i/h = 1$, or:
\begin{eqnarray}
\label{eq:deltax}
\Delta x_i=\frac{2 \pi \hbar}{p_i}\, .
\end{eqnarray}

We can express  $\Delta x_i$ as a function of $W_i$ if we suppose the relativity to hold for the virtual pairs
\begin{eqnarray}
\label{eq:dx}
\Delta x_i =\frac{2\pi \hbar c}{\sqrt{(W_i/2)^2-(W_i^0)^2}} = \frac{{2\pi\lambda_C}_i}{\sqrt{K_W^2-1}} ,
\end{eqnarray}
where ${\lambda_C}_i$ is the Compton length associated to fermion $i$.

We write the density as
\begin{eqnarray}
\label{eq:density}
N_i \approx \frac{1}{\Delta x_i^3} = \left(\frac{\sqrt{K_W^2-1}}{{2\pi\lambda_C}_i}\right)^3 .
\end{eqnarray}

Each pair can only be produced in two fermion-antifermion spin combinations: up-down and down-up. We define $N_i$ as the density of pairs for a given spin combination. It is very sensitive to $K_W$, being zero for pairs having no internal kinetic energy.

\textbullet \ The separation between the fermion and the antifermion in a pair is noted $\delta_i$. This parameter has to do with the physics of the virtual pairs. We assume it does not depend upon the fermion momentum. We will use the Compton wavelength of the fermion $\lambda_{C_i}$  as this scale:
\begin{eqnarray}
\label{eq:compton-length}
\delta_i \approx {{\lambda}_C}_i  .
\end{eqnarray}

\textbullet \ The interaction of a real photon with a virtual pair must not exchange energy or momentum with the vacuum. For instance, Compton scattering is not possible. To estimate this interaction probability, we start from the Thomson cross-section $\sigma_{Thomson}= {8 \pi}/{3}\ \alpha^2 {\lambda^2_{C}}_i$ which describes the interaction of a photon with a free electron. The factor $\alpha^2$ corresponds to the probability $\alpha$ that the photon is temporarily absorbed by the real electron times the probability $\alpha$ that the real electron releases the photon. However, in the case of the interaction of a photon with a virtual pair, the second $\alpha$ factor must be ignored since the photon is released with a probability equal to 1 as soon as the virtual pair disappears. Therefore the cross-section $\sigma$ for a real photon to interact and to be trapped by a virtual pair of fermions will be expressed as
\begin{eqnarray}
\label{eq:sigma}
\sigma_i \approx  \left(\frac{8 \pi}{3} \alpha\ Q_i^2 {\lambda^2_C}_i\right)\times 2  .
\end{eqnarray}
The photon interacts equally with the fermion and the antifermion, which explains the factor $2$.
A photon of helicity $1$ ($-1$ respectively) can interact only with a fermion or an antifermion with helicity $-1/2$ ($+1/2$ respectively) to flip temporarily its spin to helicity $+1/2$  ($-1/2$ respectively). During such a photon capture by a pair,
both fermions are in the same helicity state and cannot couple to another incoming photon in the same helicity state as the first one.

%
%

\section{Derivation of the light velocity in vacuum}
\label{sec:speedoflight}

We propose in this section a mechanism which leads to a \textbf{finite} speed of light. 
The propagation of the photon in vacuum is modeled as a series of interactions with the virtual fermions or antifermions present in the pairs. 
When a real photon propagates inside the vacuum, it interacts and is temporarily captured by a virtual pair during a time of the order of the  lifetime $\tau_i$ of the virtual pair. 
As soon as the virtual pair disappears, it releases the photon to its initial energy and momentum state.
The photon continues to propagate with a \textbf{bare} velocity $c_0$ which is assumed to be much greater than $c$. Then it interacts again with a virtual pair and so on. 
The delay on the photon propagation produced by these successive interactions implies that the velocity of light is finite.

The mean free path of the photon between two successive interactions with a $i-$type pair is:
\begin{eqnarray}
\label{eq:freepath}
\Lambda_i = \frac{1}{\sigma_i N_i}\ ,
\end{eqnarray}
where $\sigma_i$ is the cross-section for the photon capture by the virtual $i-$type pair and $N_i$ is the numerical density of virtual $i-$type pairs. 

Travelling a distance $L$ in vacuum leads on average to $N_{stop,i}$ interactions on the $i-$kind pairs. One has:
\begin{eqnarray}
\label{eq:Nstop}
N_{stop,i} = \frac{L}{\Lambda} = L{\sigma_i N_i}\ .
\end{eqnarray}

Each kind of fermion pair contributes in reducing the speed of the photon. So, if the mean photon stop time on a $i-$type pair is $\tau_i$, the mean time $\overline{T}$ for a photon to cross a length $L$ is  assumed to be:
\begin{eqnarray}
\label{eq:Tbar}
\overline{T} = L/c_0 +  \sum_{i}{N_{stop,i}\tau_i}\ .
\end{eqnarray}

The bare velocity $c_0$ is the velocity of light in an \textbf{empty} vacuum with no virtual particles. We assume that $c_0$ is infinite, which is equivalent to say that the time does not flow in an empty vacuum (with a null zero point energy). There are no "natural" time or distance scales in such an empty vacuum, whereas the $\tau_i$ and $\Delta x_i$ scales allow to build a speed scale.
So, the total delay reduces to:
\begin{eqnarray}
\label{eq:Tbar2}
\overline{T} = \sum_{i}{N_{stop,i}\tau_i}\  .
\end{eqnarray}
So, a photon, although propagating at the speed of light, is at any time resting on one fermion pair.

Using Eq. (\ref{eq:Nstop}), we obtain the photon velocity $\tilde{c}$ as a function of three parameters of the vacuum model: 
\begin{eqnarray}
\label{eq:c-1}
\tilde{c} = \frac{L}{\overline{T} }= \frac{1}{\sum_{i}{\sigma_i N_i \tau_i}} .
\end{eqnarray}

We notice that the cross-section $\sigma_i$ in (\ref{eq:sigma}) does not depend upon the energy of the photon. It implies that the vacuum is not dispersive as it is experimentally observed. 
Using Eq. (\ref{eq:tau}), (\ref{eq:density})  and (\ref{eq:sigma}), we get the final expression:
\begin{eqnarray}
\label{eq:c-2}
\tilde{c} = \frac{K_W}{\left(K_W^2-1\right)^{3/2}}\ \frac{\ 6\pi^2}{\alpha \hbar \sum_{i}{Q_i^2/({\lambda_C}_iW_i^0)}} .
\end{eqnarray}

${\lambda_C}_iW_i^0/\hbar$ is equal to the speed of light:
\begin{eqnarray}
\label{eq:lcwi}
{{\lambda_C}_iW_i^0}/{\hbar}=\frac{\hbar}{m_i c}\ m_i c^2 \frac{1}{\hbar} = c .
\end{eqnarray}
So
\begin{eqnarray}
\label{eq:c-3}
\tilde{c} = \frac{K_W}{\left(K_W^2-1\right)^{3/2}}\ \frac{\ 6\pi^2}{\alpha \sum_{i}{Q_i^2}}\ c .
\end{eqnarray}

The photon velocity depends only on the electrical charge units $Q_i$ of the virtual charged fermions present in vacuum. It depends neither upon their masses, nor upon the vacuum energy density.

The sum in Eq. (\ref{eq:c-3}) is taken over all pair types. 
Within a generation the absolute values of the electric charges are 1, 2/3 and 1/3 in units of the positron charge.
Thus for one generation  the sum writes $(1+3 \times(4/9+1/9))$. The factor 3 is the number of colours.
Each generation contributes equally, hence for the three families of the standard model:
\begin{eqnarray}
\label{eq:sommeq2}
\sum_{i}{Q_i^2} = 8 .
\end{eqnarray}

One obtains 
\begin{eqnarray}
\label{eq:c-4}
\tilde{c} = \frac{K_W}{(K_W^2-1)^{3/2}}\frac{3\pi^2} {4 \alpha}\ c .
\end{eqnarray}

The calculated light velocity $\tilde{c}$ is equal to the observed value $c$ when
\begin{eqnarray}
\label{eq:cadoublev}
\frac{K_W}{(K_W^2-1)^{3/2}}=\frac{4\alpha}{3\pi^2} ,
\end{eqnarray}
which is obtained for $K_W \approx 31.9\,$, greater than one as required.

The average speed of the photon in our medium being $c$, the photon propagates, on average, along the light cone. As such, the effective average speed of the photon is independent of the inertial frame as demanded by relativity. This mechanism relies on the notion of an absolute frame for the vacuum at rest. 
It  satisfies special relativity only in the Lorentz-Poincar\'e sense.

%
%
\section{Derivation of the vacuum permittivity}
\label{sec:permittivity}

Consider a parallel-plate capacitor with a gas inside. When the pressure of the gas decreases the capacitance decreases too until there is no more molecules in between the plates. The strange thing is that the capacitance is not zero when we hit the vacuum. In fact the capacitance has a very sizeable value as if the vacuum were a usual material body. The dielectric constant of a medium is coming from the existence of opposite electric charges that can be separated under the influence of an applied electric field $\vec{E}$. Furthermore the opposite charges separation stays finite because they are bound in a molecule. These opposite translations result in opposite charges appearing on the dielectric surfaces in regard of the metallic plates. This leads to a decrease of the effective charge, which implies a decrease of the voltage across the dielectric slab and finally to an increase of the capacitance. In our model of the vacuum the virtual pairs are the pairs of opposite charges and the separation stays finite because the electric field acts only during the  lifetime of the pairs. In an absolute \textbf{empty} vacuum the induced charges would be null because there would be no charges to be separated and the capacitance of our parallel-plate capacitor would go to zero when we would remove all molecules of the gas. We will see in this section that introducing our vacuum filled by virtual fermions will cause its electric charges to be separated and to appear at the level of $5.10^7$ electron charges per $m^2$ under an electric stress $E = 1\ V/m$.

We assume that every fermion-antifermion virtual pair of the $i$-kind bears a mean electric dipole $d_i$ given by:
\begin{eqnarray}
\label{eq:elecdipole}
\vec{d_i} = Q_i e \vec{\delta_i} .
\end{eqnarray}
where $\delta_i$ is the average size of the pairs. If no external electric field is present, the dipoles point randomly in any direction and their resulting average field is zero. We propose to give a physical interpretation of the observed vacuum permittivity  $\epsilon_0$ as originating from the mean polarization of these virtual fermions pairs in presence of an external electric field $\vec{E}$. This polarization would show up due to the dipole  lifetime dependence on the electrostatic coupling energy of the dipole to the field. In a field homogeneous at the $\delta_i$ scale, this energy is $d_i E \cos \theta$ where $\theta$  is the angle between the virtual dipole and the electric field $\vec{E}$. The electric field modifies the pair  lifetimes according to their orientation:
\begin{eqnarray}
\label{eq:taudipel}
\tau_i(\theta)= \frac{\hbar/2} {W_i - d_i E \cos \theta} .
\end{eqnarray}

Since it costs less energy to produce such an elementary dipole aligned with the field, this configuration lasts a bit longer than the others, leading to an average dipole different from zero. This average dipole $\langle D_i \rangle$ is aligned with the electric field $\vec{E}$. Its value is obtained by integration over $\theta$ with a weight proportional to the pair lifetime:
\begin{eqnarray}
\label{eq:D}
\langle D_i \rangle = \frac{\int_0^{\pi} d_i\ \cos\theta\  \tau_i(\theta)\ 2\pi \sin\theta\ d\theta}{\int_0^{\pi} \tau_i(\theta)\ 2\pi \sin\theta\ d\theta} .
\end{eqnarray}

To first order in $E$, one gets: 
\begin{eqnarray}
\label{eq:polar}
\langle D_i \rangle = d_i \frac{d_i E}{3 W_i} = {{ d_i^2}\over {3W_i}} E .
\end{eqnarray}

We estimate the permittivity $\tilde{\epsilon}_{0,i}$ due to $i$-type fermions using the relation
\newline $P_i=\tilde{\epsilon}_{0,i}E$,
where the polarization $P_i$ is equal to the dipole density
$P_i=2 N_i \langle D_i \rangle$, since the two spin combinations contribute. Thus:
\begin{eqnarray}
\label{eq:epsi}
\tilde{\epsilon}_{0,i} =2 N_i \frac{\langle D_i \rangle}{E} =2 N_i \frac{d_i^2}{3W_i} =2 N_i e^2 \frac{Q_i^2 \delta_i^2}{3W_i} .
\end{eqnarray}

Each species of fermions increases the induced polarization and therefore the vacuum permittivity. By summing over all pair species, one gets the estimation of the vacuum permittivity:
\begin{eqnarray}
\label{eq:epsi0}
\tilde{\epsilon}_{0} = e^2 \sum_{i}{2 N_i Q_i^2 \frac{\delta_i^2}{3W_i}} .
\end{eqnarray}

We can write that permittivity as a function of our units, using Eq.  (\ref{eq:energy}), (\ref{eq:density}), (\ref{eq:compton-length}) and (\ref{eq:lcwi}):
\begin{eqnarray}
\label{eq:epsi0bis}
\tilde{\epsilon}_{0} = \frac{(K_W^2-1)^{3/2}}{K_W}\frac{e^2}{24 \pi^3\hbar c}  \sum_{i}{Q_i^2} .
\end{eqnarray}

The sum is again taken over all pair types. From Eq. (\ref{eq:sommeq2}) one gets:
\begin{eqnarray}
\label{eq:permittivity}
\tilde{\epsilon}_0 = \frac{(K_W^2-1)^{3/2}}{K_W} \frac{e^2}{3\pi^3 \hbar c}\, .
\end{eqnarray}

And, from Eq. (\ref{eq:cadoublev}) one gets:
\begin{eqnarray}
\label{eq:permittivity}
\tilde{\epsilon}_0 = {\left(\frac{4\alpha}{3\pi^2}\right)}^{-1} \frac{e^2}{3\pi^3 \hbar c}=\frac{e^2}{4\pi\hbar c\alpha} =8.85\, 10^{-12} F/m
\end{eqnarray}

It is remarkable that Eq. (\ref{eq:cadoublev}) obtained from the derivation of the speed of light leads to a calculated permittivity $\tilde{\epsilon}_0$ exactly equal to  the observed value of $\epsilon_0$. 


%
%

\section{Derivation of the vacuum permeability}
\label{sec:permeability}
The vacuum acts as a highly paramagnetic substance. When a torus of a material is energized through a winding carrying a current $I$, there is a resulting magnetic flux density $B$ which is expressed as:  
\begin{eqnarray}
\label{eq:mu-1}
B = \mu_0 n I + \mu_0 M .
\end{eqnarray}
where $n$ is the number of turns per unit of length, $nI$ is the magnetic intensity in $A/m$ and $M$ is the corresponding magnetization induced in the material and is the sum of the induced magnetic moments divided by the corresponding volume. 
In an experiment where the current $I$ is kept a constant and where we lower the quantity of matter in the torus, $B$ decreases. As we remove all matter, $B$ gets to a non zero value: $B = \mu_0 n I$ showing experimentally that the vacuum is paramagnetic with a vacuum permeability $\mu_0 = 4\pi\ 10^{-7} {N/A^2}$.

We propose to give a physical interpretation to the observed vacuum permeability as originating from the magnetization of the charged virtual fermions pairs under a magnetic stress,  following the same procedure as in the former section.

Each charged virtual fermion carries a magnetic moment proportional to the Bohr magneton:
\begin{eqnarray}
\label{eq:magneton}
\mu_i = \frac{eQ_i\ c{\lambda_C}_i}{2} .
\end{eqnarray}

Since the total spin of the pair is zero, and since fermion and antifermion have opposite charges, each pair carries twice the magnetic moment of one fermion. The coupling energy of a $i$-kind pair to an external magnetic field $\vec{B}$ is then $-2 \mu_i B \cos \theta$ where $\theta$ is the angle between the magnetic moment and the magnetic field $\vec{B}$. 
The pair  lifetime is therefore a function of the orientation of its magnetic moment with respect to the applied magnetic field:
\begin{eqnarray}
\label{eq:taumag}
\tau_i(\theta)= \frac{\hbar/2}{W_i - 2 \mu_i B \cos \theta} .
\end{eqnarray}

As in the electrostatic case, pairs with a dipole moment aligned with the field last a bit longer than anti-aligned pairs. 
This leads to a non zero average magnetic moment $<\mathcal{M}_i>$ for the pair, aligned with the field and given, to first order in $B$, by:
\begin{eqnarray}
\label{eq:magnet}
<\mathcal{M}_i> = \frac{4\mu_i^2}{3W_i} B .
\end{eqnarray}

The  volume magnetic moment is
$M_i  = {2 N_i <\mathcal{M}_i>}$, since one takes into account the two spin states per cell.

The contribution $\tilde{\mu}_{0,i}$ of the $i$-type fermions to the vacuum permeability is given by $ B=\tilde{\mu}_{0,i}M_i $ or
${1}/{\tilde{\mu}_{0,i}}={M_i}/{B}$.

This leads to the estimation of the vacuum permeability
\begin{eqnarray}
\label{eq:permeability-1}
\frac{1}{\tilde{\mu}_0}=\sum_{i}{\frac{M_i}{B}} = \sum_{i}{\frac{8 N_i\mu_i^2}{3W_i}}= c^2 e^2\sum_{i}{\frac{2 N_iQ_i^2{\lambda^2_C}_i}{3W_i}} .
\end{eqnarray}

Using Eq. (\ref{eq:energy}), (\ref{eq:density}) and (\ref{eq:lcwi}) and summing over all pair types, one obtains
\begin{eqnarray}
\label{eq:permeability-3}
\tilde{\mu}_0 = \frac{K_W}{(K_W^2-1)^{3/2}} \frac{24\pi^3\hbar}{c\,e^2 \sum_{i}{Q_i^2}}= \frac{K_W}{(K_W^2-1)^{3/2}} \frac{3\pi^3 \hbar}{ c\,e^2}\, .
\end{eqnarray}

Using the $K_W$ value constrained by the calculus of $c$  (\ref{eq:cadoublev}), we end up with:
\begin{eqnarray}
\label{eq:permeability-4}
\tilde{\mu}_0 = \frac{4\pi\alpha}{3} \frac{3\ \hbar}{c\,e^2} =\frac{4\pi\alpha\hbar}{c\,e^2} = 4\pi 10^{-7}N/A^2 .
\end{eqnarray}

It is again remarkable that Eq. (\ref{eq:cadoublev}) obtained from the derivation of the speed of light leads to a calculated permeability $\tilde{\mu}_0$ equal to the right $\mu_0$ value. 

We notice that the permeability and the permittivity do not depend upon the masses of the fermions, as in Ref.~~\cite{Leuchs}.
The electric charges and the number of species are the only important parameters. 
This is at variance with the common idea that the energy density of the vacuum is the dominant factor~\cite{Latorre}.

This expression, combined with the expression~(\ref{eq:permittivity}) of the calculated permittivity, verifies the Maxwell relation, typical of wave propagation, $ \tilde{\epsilon}_0 \tilde{\mu}_0  = 1/c^2$,
although our mechanism for a finite $c$ is purely corpuscular.

%
%

\section{A generalized model }

We have shown that our model of vacuum leads to coherent calculated values of $c$, $\epsilon_0$ and $\mu_0$ equal to the observed values if we assume that the virtual fermion pairs are produced with an average energy which is about 30 times their rest mass.

This solution corresponds to some \textbf{natural} hypotheses for the density and the size of the virtual pair, the cross-section with real photons and their capture time.
We can generalize the model by introducing the free parameters $K_N$, $K_{\delta}$, $K_{\sigma}$ and $K_\tau$ in the expressions~(\ref{eq:density}), (\ref{eq:compton-length}), (\ref{eq:sigma}) and (\ref{eq:Tbar}) of the physical quantities:
\begin{eqnarray}
\label{eq:1}
N_i = K_N \left( \frac{\sqrt{K_W^2-1}}{2\pi{\lambda_C}_i} \right)^3
\end{eqnarray}
\begin{eqnarray}
\label{eq:2}
\delta_i = K_{\delta}\, {\lambda_C}_i
\end{eqnarray}
\begin{eqnarray}
\label{eq:3}
\sigma_i = K_{\sigma} \frac{16 \pi}{3} \alpha\, Q_i^2 {\lambda^2_C}_i
\end{eqnarray}
\begin{eqnarray}
\label{eq:4}
\overline{T} = \sum_{i}{N_{stop,i}K_\tau\tau_i}\ .
\end{eqnarray}
$K_N$, $K_{\delta}$, $K_{\sigma}$ and $K_\tau$ are assumed to be universal factors independent of the fermion species. Their values are expected to stay close to $1$. Among the model parameters, $K_W$ is the only unconstrained unknown.

The general solutions of the calculation of $c$ (\ref{eq:c-4}), $\epsilon_0$ (\ref{eq:permittivity}) and $\mu_0$ (\ref{eq:permeability-4}) read now:
\begin{eqnarray}
\label{eq:5}
\tilde{c} = \frac{1}{K_N K_{\sigma}K_\tau } \frac{K_W}{(K_W^2-1)^{3/2}}\frac{3\pi^2} {4 \alpha}\ c\, ,
\end{eqnarray}
\begin{eqnarray}
\label{eq:6}
\tilde{\epsilon}_0 = K_N K_{\delta}^2 \frac{(K_W^2-1)^{3/2}}{K_W} \frac{e^2}{3\pi^3 \hbar c}\, ,
\end{eqnarray}
\begin{eqnarray}
\label{eq:7}
\tilde{\mu}_0 = \frac{1}{K_N} \frac{K_W}{(K_W^2-1)^{3/2}} \frac{3\pi^3 \hbar}{c\,e^2}\, ,
\end{eqnarray}
from which one can get, for instance:
\begin{itemize}
\item  first $\tilde{\epsilon}_0\tilde{\mu}_0=K_{\delta}^2/c^2$ which implies $K_{\delta} = 1$\ ,
\item  then either $\tilde{\epsilon}_0$ or $\tilde{\mu}_0$ fixes $\frac{1}{K_N} \frac{K_W}{(K_W^2-1)^{3/2}} = \frac{4 \alpha}{3\pi^2}\ $
\item  which applied to $\tilde{c}$ gives $K_{\sigma}K_\tau = 1$.
\end{itemize}
So,  the model parameters satisfy:
\begin{eqnarray}
\label{eq:9}
K_{\delta} = 1\ , K_{\sigma}K_\tau = 1\ ,
\frac{1}{K_N} \frac{K_W}{(K_W^2-1)^{3/2}} = \frac{4\alpha}{3\pi^2}\ .
\end{eqnarray}

$K_{\delta}$ is precisely constrained to its first guess value. More relations or observables are required to extract the other quantities and check this vacuum model. A measurement of the expected fluctuations of the speed of light, and a measurement of speed of light variations with light intensity, as discussed in the following sections, could bring such relations.

%
%

\section{Transit time fluctuations}
\label{sec:prediction transit}
\subsection{Prediction}

Quantum gravity theories including stochastic fluctuations of the metric of compactified dimensions, predict a fluctuation $\sigma_t$ of the propagation time of photons~\cite{Yu-Ford}. However observable effects are expected to be too small to be experimentally tested. 
It has been also recently predicted that the non commutative geometry at the Planck scale should produce a spatially coherent space-time jitter~\cite{Hogan}.

In our model we also expect fluctuations of the speed of light $c$. Indeed in the mechanism proposed here $c$ is due to the effect of successive interactions and transient captures of the photon with the virtual particles in the vacuum. Thus statistical fluctuations of $c$ are expected, due to the statistical fluctuations of the number of interactions $N_{stop}$ of the photon with the virtual pairs and to the capture time fluctuations.

The propagation time of a photon which crosses a distance $L$ of vacuum is
\begin{eqnarray}
\label{eq:transit}
t = \sum_{i,k}{ t_{i,k}} ,
\end{eqnarray}
where $t_{i,k}$ is the duration of the $k^{th}$ interaction on an $i$-kind pair. As in section \ref{sec:speedoflight}, let $ N_{stop,i}$ be the mean number of such interactions.
The variance of $t$ due to the statistical fluctuations of $N_{stop,i}$ is:
\begin{eqnarray}
\label{eq:sig1}
\sigma_{t,N}^2 = \sum_{i}{N_{stop,i} K_\tau^2\tau_i^2} .
\end{eqnarray}
The photon may arrive on a virtual pair any time between its birth and its death. If we assume a flat probability distribution between $0$ and $\tau_i$, the mean value of $t_{i,k}$ is $\tau_i/2$, so one has $K_\tau=1/2$. The variance of the stop time is $(K_\tau\tau_i)^2/3$:
\begin{eqnarray}
\label{eq:sig2}
\sigma_{t,\tau}^2 = \sum_{i}{N_{stop,i} \frac{(K_\tau\tau_i)^2}{3}} .
\end{eqnarray}
Then
\begin{eqnarray}
\label{eq:fluctu}
\sigma_t^2 = \sum_{i}{N_{stop,i} (K_\tau\tau_i)^2(1+\frac{1}{3}})= \frac{4\,K_\tau^2}{3}\sum_{i}{N_{stop,i} \tau_i^2}.
\end{eqnarray}
And, using Eq. (\ref{eq:Nstop}):
\begin{eqnarray}
\label{eq:fluctuation-0}
\sigma_t^2 = \frac{4 K_\tau^2L}{3} \sum_{i}{\sigma_i N_i \tau_i^2} . 
\end{eqnarray}
Once reduced, the current term of the sum is proportional to ${\lambda_C}_i$. Therefore the fluctuations of the propagation time are dominated by virtual $e^+e^-$ pairs. Neglecting the other fermion species, and using $\sigma_e N_eK_\tau\tau_e=1/(8c)$, one gets :
\begin{eqnarray}
\label{eq:formulesigmat2}
\sigma_t^2 = \frac{K_\tau\tau_eL}{6c}= \frac{K_\tau{\lambda_C}_eL}{24 K_W c^2} .
\end{eqnarray}
So
\begin{eqnarray}
\label{eq:fluctuation}
{\sigma_t} = \sqrt{\frac{L}{c}}\sqrt{\frac{{\lambda_C}_e}{c}}\sqrt{\frac{K_\tau}{K_W}}\frac{1}{\sqrt{24}}  .
\end{eqnarray}

For the simple solution of the vacuum model where $K_W=31.9$\, and $K_\tau=1/2$ the predicted fluctuation is:
\begin{eqnarray}
\label{eq:fluctuation-2}
\sigma_t \approx 53 \ as\ m^{-1/2} \sqrt{L(m)} .
\end{eqnarray}

This corresponds for instance on a $1\ m$ long travel, to an average of $8.\, 10^{13} $ stops by $e^+e^-$ pairs during which the photon stays on average $5\, 10^{-24} s$ (it spends $7/8$ of its time trapped on other species pairs).
Fluctuations of both quantities lead to this $50\ as$ expected dispersion on the photon transit time which represents a $1.5\,10^{-8}$ relative fluctuation over a meter.

This prediction must be modulated by the remaining degree of freedom on $K_N$ or $K_W$, but the mechanism would loose its physical basis if $\sigma_t$ would not have that order of magnitude.

A positive measurement of  $\sigma_t$, apart from being a true revolution, would tighten our understanding of the fundamental constants in the vacuum, by fixing the ratio $K_\tau/K_W$.

The experimental way to test fluctuations is to measure a possible time broadening of a light pulse travelling a distance $L$ of vacuum. This may be done using observations of brief astrophysical events, or dedicated laboratory experiments. 

\subsection{Constraints from astrophysical observations}
The very bright GRB 090510, detected by the Fermi Gamma-ray Space Telescope~\cite{Abdo}, at MeV and GeV energy scale,
presents short spikes in the $8~keV - 5~MeV$ energy range, with the narrowest widths of the order of $10\,ms$. Observation of the optical after glow, a few days later by ground based spectroscopic telescopes gives a common redshift of $z = 0.9$. This corresponds to a distance, using standard cosmological parameters, of about $2\, 10^{26} m$. Translated into our model, this sets a limit of about $0.7\, fs\, m^{-1/2}$ on $c$ fluctuations. It is important to notice that there is no expected dispersion of the bursts in the interstellar medium at this energy scale. 

If we move six orders of magnitude down in distances we arrive to kpc and pulsars.
Short microbursts contained in main pulses from the Crab pulsar have been recently observed at the Arecibo Observatory telescope at 5 GHz~\cite{Crab-pulsar-2010}. The frequency-dependent delay caused by dispersive propagation through the interstellar plasma is corrected using a coherent dispersion removal technique. 
The mean time width of these microbursts after dedispersion is about 1~$\mu$s, much larger than the expected broadening caused by interstellar scattering. If this unknown broadening would not be correlated to the emission properties, it could come from $c$ fluctuations of about $0.2\, fs\, m^{-1/2}$.

In these observations of the Crab pulsar, some very sporadic pulses with a duration of less than $1 ns$ have been observed at 9 GHz~\cite{Crab-pulsar-2007}.
This is 3 orders of magnitude smaller than the usual pulses. These nanoshots can occasionally be extremely intense, exceeding $2\, MJy$, and have
an unresolved duration of less than $0.4\, ns$ which corresponds to a light-travel size $c\delta t \approx 12\, cm$. From this the implied brightness temperature is $2\ 10^{41} K$.
Alternatively we might assume the emitting structure is moving outward with a Lorentz factor $\gamma_b \approx 10^2 - 10^3$. In that case, the size estimate
increases to $10^3 - 10^5\, cm$, and the brightness temperature decreases to $10^{35} - 10^{37}\, K$. 
We recall that the Compton temperature is $10^{12}\, K$ and that the Planck temperature is $10^{32}\, K$ so the phenomenon, if real,
would be way beyond known physics.
We emphasize also two features. Firstly, these nanoshots are contained in a single time bin (2~\,ns at 5~\,GHz and 0.4~\,ns at 9~\,GHz) corresponding to a time width less than $2/\sqrt{12} \approx 0.6\, ns$ at $5~\,GHz$ and $0.4/\sqrt{12} \approx 0.1\,ns$ at $9~\,GHz$, below the expected broadening caused by interstellar scattering. Secondly, their frequency distributions appear to be almost monoenergetic and very unusual, since the shorter the pulse the narrower its reconstructed energy spectrum.

\subsection{Constraints from Earth bound experiments}
The very fact that the predicted statistical fluctuations should go like the square root of the distance implies the exciting idea that experiments on Earth do compete with astrophysical constraints, since going from the kpc down to a few hundred meters, which means a distance reduction by a $10^{17}$ factor, we expect fluctuations in the $fs$ range.

An experimental setup using femtosecond lasers pulses sent to a rather long multi-pass cavity equipped with metallic mirrors could be able to detect such a phenomenon.

The attosecond laser pulse generation and characterization, by itself, might allow already to set the best limit on $\sigma_t$. This limit would be of the order of our predicted value in the simplest version of the model~\cite{atto-1}. However, for the time being, the unambiguous measurement of the pulse time spread is available only for $fs$ pulses through the correlation of the short pulses in a non linear crystal.

\section{Modification of the light speed in extremely intense light pulses}
\label{sec:vacmod}
The vacuum, considered as a peculiar medium, should be able to undergo changes. This is suspected to be the case in the Casimir effect which predicts a pressure to be present between electrically neutral conducting surfaces~\cite{Casimir}. This force has been observed in the last decade by several experiments~\cite{Casimir-exp-1} and is interpreted as arising from the modification of the zero-point energies of the vacuum due to the presence of material boundaries.

The vacuum can also be seen as the triggering actor in the spontaneous decay of excited atomic states through a virtual photon stimulating the emission~\cite{Purcell}. In that particular case experimentalists were able to change the vacuum, producing a huge increase~\cite{Goy} or a decrease~\cite{Hulet} of the spontaneous emission rate by a modification of the virtual photon density.

This model predicts that the local vacuum is also modified by a light beam because of photon capture by virtual pairs, which in a sense pumps the vacuum. 

Let us apply the mechanism exposed in section \ref{sec:speedoflight} to the propagation of a pulse  when photon densities are not negligibly small compared to the $e^+e^-$ pair density.

If the pulse is fully circularly polarized, all its photons bear the same helicity. So, a photon cached by a pair makes it transparent to the other incoming photons, till it jumps on another one.

If the photon density is $N_\gamma$, the fraction of $i-$type species masked this way is, to first order in  $N_\gamma$, equal to:
\begin{eqnarray}
\label{eq:masked}
\Delta N_i/N_i=N_\gamma K_\sigma \sigma_i c K_\tau\tau_i  .
\end{eqnarray}
So, from (\ref{eq:9})
\begin{eqnarray}
\label{eq:masked-2}
\Delta N_i=N_\gamma N_i \sigma_i c\tau_i  .
\end{eqnarray}
The remaining densities available to interact with photons are $N_i-\Delta N_i$. So the speed of light is given by :
\begin{eqnarray}
\label{eq:cstar}
\tilde{c}^* = \frac{1}{\sum_{i}{\sigma_i (N_i-\Delta N_i)\tau_i}} =\frac{1}{\sum_{i}{\sigma_i N_i
\tau_i(1-N_\gamma\sigma_i c\tau_i )}}  .
\end{eqnarray}

$\sigma_i \tau_i$ being proportional to ${{\lambda^3_C}_i}$, we keep only the $e^+e^-$ contribution in the corrective term. Using (\ref{eq:c-1}), it comes
\begin{eqnarray}
\label{eq:cstar-2}
\tilde{c}^* =\frac{ c}{1- N_\gamma N_e\sigma_e^2 c^2\tau_e^2}\,  .
\end{eqnarray}

Noticing that $N_e\sigma_e \tau_e = 1/8c$, this reduces to:
\begin{eqnarray}
\label{eq:cstar-2}
\tilde{c}^* =\frac{ c}{1-N_\gamma/(64 N_e)}\, .
\end{eqnarray}

So, one ends up with:
\begin{eqnarray}
\label{dcc}
\frac{\delta{c}}{c} =\frac{N_\gamma}{64 N_e}\, ,
\end{eqnarray}
which shows that $c$ would be an increasing function of the photon densities. This anti Kerr effect is directly related to the $e^+e^-$ pair density. One can express it as a function of $K_W$, using (\ref{eq:1}) and (\ref{eq:9}):
\begin{eqnarray}
\label{dcckw}
\frac{\delta{c}}{c} =\frac{N_\gamma\pi\alpha{{\lambda^3_C}_e}}{6K_W}\, .
\end{eqnarray}

This prediction could in principle be tested in a dedicated laboratory experiment where
one huge intensity pump pulse would be used to stress the vacuum and change the transit time of a small probe pulse going in the same direction and having the same circular polarization (or going in the opposite direction with the opposite polarization). Other helicity combinations would give no effect on the transit time.

Let us convert Eq. \ref{dcckw} into numbers. Using $K_W=31.9$, $N_e$ amounts to:
\begin{eqnarray}
\label{eq:densnum-e+e-}
N_e \approx 2\ 10^{39}\ e^+e^-/m^3 \,.
\end{eqnarray}

Now, the photon density in a pulse of power $P$, frequency $\omega$ and section $S$ is:
\begin{eqnarray}
\label{eq:densphot}
N_\gamma = \frac{P}{\hbar\omega S c} = \frac{P\lambda}{2\pi\hbar S c^2}\,.
\end{eqnarray}
A petawatt source at $\lambda\approx .5\,\mu m$, such as the one of Ref.~~\cite{Bayramian}, allows to reach focused irradiances of
the order of $P/S=10^{23}W/cm^2$ in a few $\lambda^3$ volume. This means photons densities in the range:
\begin{eqnarray}
N_\gamma = \frac{10^{27}\ .5 10^{-6}}{6.6\, 10^{-34}\ 9\, 10^{16}}= 10^{37}\ \gamma/ m^{3} \,,\label{eq:densphot}
\end{eqnarray}
leading to:
\begin{eqnarray}
\label{eq:ratio}
N_\gamma/N_e \approx  5\,10^{-3}\,,
\end{eqnarray}
and a relative effect on $c$ of $8\,10^{-5}$ over a short length. This would affect the time transit of a probe pulse focused through the same volume. Creating a $1\ fs$ advance on that probe pulse would need a common travel with such a pump pulse over a distance of $4\ mm$. This seems a difficult challenge, but we think that this matter deserves a specific study to build a real proposal for testing this prediction.

\section{Conclusions}

We describe the ground state of the unperturbed vacuum as containing a finite density of charged virtual fermions. 
Within this framework, the finite speed of light is due to successive transient captures of the photon by these virtual particles.
$\epsilon_0$ and $\mu_0$ originate also simply from the electric polarization and from the magnetization of these virtual particles when the vacuum is stressed by an electrostatic or a magnetostatic field respectively. 
Our calculated values for $c$, $\epsilon_0$ and $\mu_0$ are equal to the measured values when the virtual fermion pairs are produced with an average energy of about $30$ times their rest mass.
This model is self consistent and it proposes a quantum origin to the three electromagnetic constants. 
The propagation of a photon being a statistical process, we predict fluctuations of the speed of light. It is shown that this could be within the grasp of nowadays experimental techniques and we plan to assemble such an experiment.
Another prediction is a light propagation faster than $c$ in a high density photon beam.

\section*{Acknowledgments}

The authors thank N. Bhat, J.P. Chambaret , I. Cognard, J. Ha\"{\i}ssinski, P. Indelicato, J. Kaplan, P. Wolf and F. Zomer for fruitful discussions, and  J. Degert, E. Freysz, J. Oberl\' e and M. Tondusson for their collaboration on the experimental aspects. This work has benefited from a GRAM\footnote{CNRS INSU/INP program with CNES \& ONERA participations (Action Sp\' ecifique "Gravitation, R\'ef\'erences, Astronomie, M\'etrologie")} funding.


\newpage
\begin{thebibliography}{00}
\bibitem{lamb} W.E. Lamb and R.C. Retherford, Phys. Rev. 72 (1947) 241-243.
\bibitem{bare-charge} I. Levine et al., Phys. Rev. Lett. 78 (1997) 424-427.
\bibitem{magnetic-moment} J. Schwinger, Phys. Rev. 73 (1948) 416-417.
\bibitem{Davier} M. Davier, A. Hoecker, B. Malaescu and Z. Zhang, Eur.Phys.J. C71 (2011) 1515.
\bibitem{Kittel} Ch. Kittel, Elementary Solid State Physics, John Wiley \& Sons (1962) 120.
\bibitem{Leuchs} G. Leuchs, A.S. Villar and L.L. Sanchez-Soto, Appl. Phys. B 100 (2010) 9-13.
\bibitem{Yu-Ford} Yu H. and Ford L.H., Phys. Lett. B 496 (2000) 107-112.
\bibitem{Hogan} C. J. Hogan, ArXiv.org 1002.4880 (2011).
\bibitem{Abdo} Abdo A.A. et al., Nature 462 (2009) 331-334.
\bibitem{Crab-pulsar-2010} Crossley J.H. et al., The Astrophysical Journal 722 (2010) 1908-1920.
\bibitem{Crab-pulsar-2007} Hankins T.H. and Eilek J.A., The Astrophysical Journal 670 (2007) 693-701.
\bibitem{Casimir} H. Casimir, Phys. Rev., 73 (1948) 360.
\bibitem{Casimir-exp-1} S.K. Lamoreaux, Rep. Prog. Phys. 68 (2005) 201.
\bibitem{Purcell} E.M. Purcell, Phys. Rev. 69 (1946) 681.
\bibitem{Goy} P. Goy, J.M. Raymond, M. Gross and S. Haroche Phys. Rev. Lett. 50 (1983) 1903.
\bibitem{Hulet} R. G. Hulet, E.S. Hilfer and D. Kleppner, Phys. Rev. Lett. 55 (1985) 2137.
\bibitem{Latorre} J. I. Latorre et al., Nuclear Physics B437 (1995) 60.
\bibitem{atto-1} E. Goulielmakis et al., Science 320, 1614 (2008).
\bibitem{Bayramian} A. Bayramian et al., JOSA B 25 - Issue 7 (2008) B57-B61.
\end{thebibliography}
\end{document}